\documentclass
[aps,prl,twocolumn,showpacs,lengthcheck,preprintnumbers]{revtex4}%
\usepackage{amsfonts}
\usepackage{amsmath}
\usepackage{amssymb}
\usepackage{graphicx}%
\providecommand{\U}[1]{\protect\rule{.1in}{.1in}}
\begin{document}
\preprint{Phys. Rev. A \textbf{82}, 053606 (2010)}
\title{Three-dimensional gap solitons in Bose-Einstein condensates supported by
one-dimensional optical lattices}
\author{A. Mu\~{n}oz Mateo$^{1}$, V. Delgado$^{1}$, and Boris A. Malomed$^{2} $}
\affiliation{$^{1}$Departamento de F\'{\i}sica Fundamental II, Universidad de La Laguna,
38206 La Laguna, Tenerife, Spain}
\affiliation{$^{2}$Department of Physical Electronics, School of Electrical Engineering,
Faculty of Engineering, Tel Aviv University, Tel Aviv 69978, Israel}


\begin{abstract}
We study fundamental and compound gap solitons (GSs) of matter waves in
one-dimensional (1D) optical lattices (OLs) in a three-dimensional (3D)
weak-radial-confinement regime, which corresponds to realistic experimental
conditions in Bose-Einstein condensates (BECs). In this regime GSs exhibit
nontrivial radial structures. Associated with each 3D linear spectral band
exists a family of fundamental gap solitons that share a similar transverse
structure with the Bloch waves of the corresponding linear band. GSs with
embedded vorticity $m$ may exist \emph{inside} bands corresponding to other
values of $m$. Stable GSs, both fundamental and compound ones (including
vortex solitons), are those which originate from the bands with lowest axial
and radial quantum numbers. These findings suggest a scenario for the
experimental generation of robust GSs in 3D settings.

\end{abstract}
\date{11 June 2010}
\maketitle

\section{I. INTRODUCTION}

A ubiquitous tool for the control of collective excitations in Bose-Einstein
condensates (BECs) is provided by optical lattices (OLs), which are induced by
the interference of laser beams illuminating the condensate \cite{mor1}. OLs
are especially efficient in supporting matter-wave solitons. It has been
predicted that two- and three-dimensional (2D and 3D) OLs can stabilize
solitons against collapse \cite{BBB1}. OLs acting in the combination with
repulsive interactions can give rise to diverse species of gap solitons (GSs),
in one dimensional (1D) \cite{1D,sub} and multidimensional
\cite{Konotop,vortex} geometries. Quasistable GSs were created in the $^{87}%
$Rb condensate loaded into a cigar-shaped trap incorporating an axial OL
\cite{ober1}. Extended states, built as segments of nonlinear Bloch waves
trapped in the OL, were also reported \cite{gap-wave-experiment}.

Most theoretical studies of GSs have been carried out in the quasi-1D regime,
assuming that the transverse confinement is tight enough to reduce the wave
function in the transverse plane to the ground state of the corresponding 2D
harmonic oscillator (HO) \cite{Gora,Luca}. In this case, the description of
the relevant dynamics amounts to the 1D Gross-Pitaevskii equation (GPE) in the
axial direction, which follows from the underlying 3D equation after averaging
out the radial (transverse) degree of freedom. Since in the quasi-1D regime
the radial quantum $\hbar\omega_{\bot}$ is the largest energy scale of the
system, GSs cannot decay by exciting higher-order radial modes, being
therefore particularly stable. However, the creation of genuine quasi-1D
settings, strongly squeezed in the radial direction ---for instance, those
realizing the Tonks-Girardeau gas \cite{TG}--- is a challenging experimental
problem. The settings used in experiments with GSs
\cite{mor1,ober1,gap-wave-experiment} actually correspond to the weak
transverse confinement (see details in Sec. IV).

In this work we aim to predict matter-wave GSs in the regime of the weak
radial confinement, characterized by a recoil energy of the axial OL, $E_{R}$,
comparable to $\hbar\omega_{\bot}$. While this regime is most relevant to the
experiment, the formation of GSs under these conditions was not yet studied
theoretically. For instance, $E_{R}/\hbar\omega_{\bot}=1$ corresponds to the
$^{87}$Rb condensate, with \textsl{s}-wave scattering length $a_{s}=5.29$ nm,
confined by the combination of the transverse trapping frequency $\omega
_{\bot}/2\pi=240$ Hz and axial OL of period $d=1.55$ $\mathrm{\mu}$m (physical
results given in this article correspond to this typical setting). In this
regime, the axial GS structure may readily excite higher modes of the radial
confinement; hence the 3D character of the dynamics is essential and the 1D
reduction cannot be used. The situation is somewhat similar to that for
quasi-1D GSs, which were predicted, in the framework of the density-functional
description, in fermionic superfluids \cite{SKA}. In that case, the underlying
Fermi distribution implies the filling of many transverse energy levels.

While there are models that generalize the 1D GPE by taking into account small
deviations from the one-dimensionality \cite{Gora,Luca}, we consider the
setting in which the axial and radial directions are equally important and
inseparable; hence the use of the full 3D equation is necessary. We
demonstrate that stable solitons, which are true gap modes in terms of the
underlying 3D band-gap structure, exist in this regime, suggesting
possibilities for the creation of robust 3D solitons. This objective is of
principal significance because, thus far, no truly 2D or 3D matter-wave
solitons, nor their counterparts in optical media with the Kerr nonlinearity,
have been created, in spite of many theoretical predictions \cite{review}. We
also find solitons \emph{inside} the bands, which may exist due to the
difference in the azimuthal index between the soliton and the band.

\section{II. THREE-DIMENSIONAL GAP SOLITONS}

The 3D GPE, with the radial-HO and axial potentials, $V_{\bot}(\mathbf{r}%
_{\bot})=(M/2)\omega_{\bot}^{2}r_{\bot}^{2}$ and $V_{z}(z)=sE_{R}\sin
^{2}\left(  \pi z/d\right)  $, is%
\begin{equation}
i\hbar\psi_{t}\!=\!\left[  -\!\left(  \hbar^{2}/2M\right)  \!\nabla
^{2}\!+V_{\bot}(\mathbf{r}_{\bot})+V_{z}(z)+gN\!\left\vert \psi\right\vert
^{2}\right]  \!\psi,\label{eq1}%
\end{equation}
where $N$ and $M$ are the atomic number and mass, $g\equiv4\pi\hbar^{2}%
a_{s}/M$, $d$ and $s$ are the period and depth of the OL, the recoil energy is
$E_{R}=\left(  \pi\hbar\right)  ^{2}/\left(  2Md^{2}\right)  $, and the norm
of the wave function is $1$ \cite{RCG}.

\begin{figure}[ptb]
\begin{center}
\includegraphics[
width=8.6cm
]{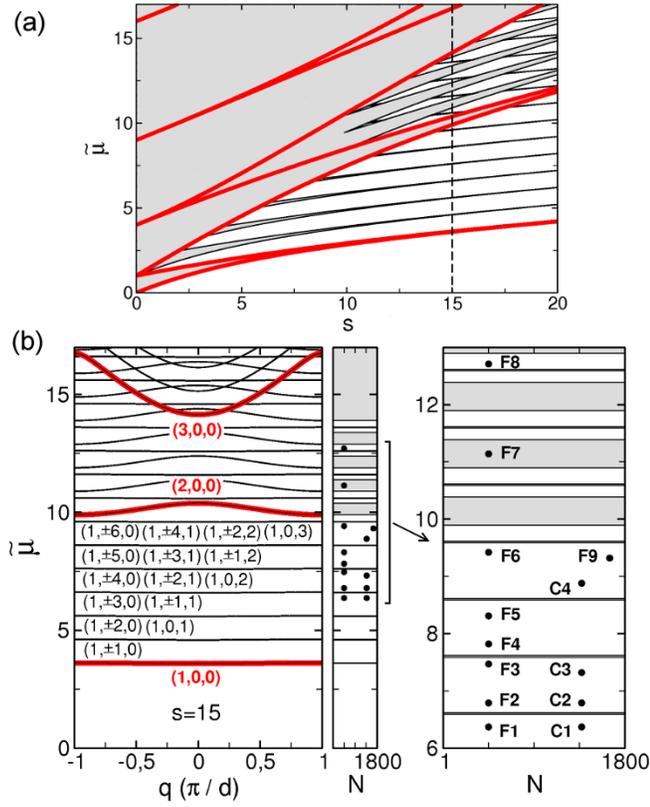}
\end{center}
\caption{(Color online) The band-gap structure, produced by the linearized
GPE, with equal axial and transverse energies, $E_{R}/\hbar\omega_{\bot}=1$
(white areas are gaps). The normalized chemical potential is shown vs the
scaled lattice depth, $s$ (a), and as a function of quasimomentum $q$ for
$s=15$ (b). The right-hand panels in (b) indicate the location of the GSs
displayed in this paper. Sets $(n,m,n_{r})$ represent the quantum numbers, and
the bold red curves depict the band-gap structure in the 1D model.}%
\label{Fig1}%
\end{figure}

Figure \ref{Fig1}(a) displays the scaled shifted chemical potential,
$\widetilde{\mu}\equiv(\mu-\hbar\omega_{\bot})/E_{R}$, of the noninteracting
3D condensate ($g=0$) with $E_{R}/\hbar\omega_{\bot}=1$, as a function of the
OL strength $s$. In the plot, bandgaps separate shaded Bloch bands. For the
sake of comparison, the band-gap diagram obtained from the corresponding 1D
equation is also shown (bold red lines). Since the energy levels of the radial
HO are $E=(2n_{r}+|m|+1)\hbar\omega_{\bot}$, where $n_{r}=0,1,2,\ldots$ and
$m=0,\pm1,\pm2,\ldots$ are the radial and azimuthal quantum numbers, the
excitation of transverse modes gives rise to a series of replicas of 1D Bloch
bands, shifted up in the energy by multiples of $\hbar\omega_{\bot}/E_{R}$. In
what follows, we present most of our results for a strong OL with\textbf{\ }%
$s=15$, which corresponds to the dashed vertical line in Fig. \ref{Fig1}(a).
The respective spectrum is shown in Fig. \ref{Fig1}(b), which displays
$\tilde{\mu}$ as a function of quasimomentum $q$ in the first Brillouin zone.
The 3D bands are characterized by quantum-number sets $(n,m,n_{r})$, where
$n=1,2,3,\ldots$ is the band index of the corresponding 1D axial problem. As
in Fig. \ref{Fig1}(a), the superimposed bold red lines represent the results
generated by the corresponding 1D equation.
As indicated in Fig. \ref{Fig1}(b),\ the lowest band corresponds to
$(n,m,n_{r})=(1,0,0)$. The next two bands, with numbers $(1,\pm1,0)$, have
equal chemical potentials, being replicas of the lowest band shifted upward by
$\hbar\omega_{\bot}/E_{R}$. Likewise, the next three bands, with numbers
$(1,\pm2,0)$ and $(1,0,1)$, are shifted by $2\hbar\omega_{\bot}/E_{R}$ and so
on. Dots in the right-hand panels in Fig. \ref{Fig1}(b) indicate the location
of the GSs that we consider in this paper, with the horizontal axes indicating
the number of $^{87}$Rb atoms in each nonlinear state.

\begin{figure}[ptb]
\begin{center}
\includegraphics[
width=8.6cm
]{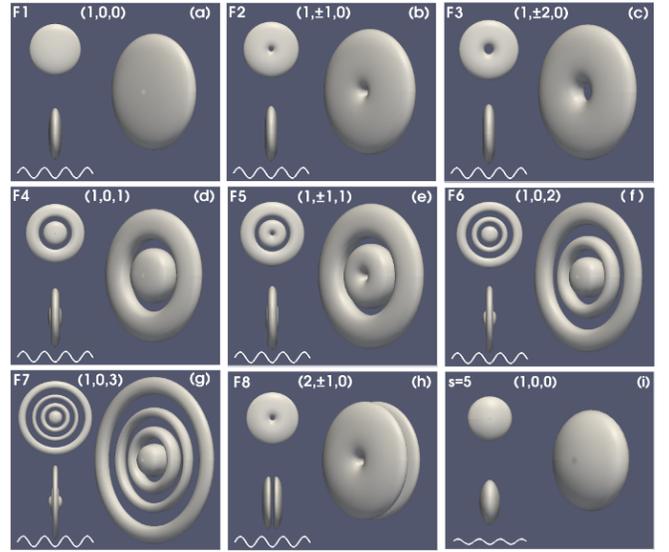}
\end{center}
\caption{(Color online) Isosurfaces of the atomic density, taken at $10\%$ of
the maximum value, for GSs corresponding to points F1--F8 in Fig.
\ref{Fig1}(b). Panel (i) depicts a GS in a weaker lattice, with $s=5$. The
respective quantum-number sets, $(n,m,n_{r})$, are indicated in each panel.
The length scale of the left part in each panel (reduced to $45\%$ in
comparison with the image on the right) is gauged by the OL period.}%
\label{Fig2}%
\end{figure}

GS solutions have been obtained as numerical solutions of the stationary
version of Eq. (\ref{eq1}), using the Newton continuation method with the
Laguerre-Fourier functional basis. In Fig. \ref{Fig2} we display a set of
solutions for GSs which correspond to points F1--F8 in Fig. \ref{Fig1}(b).
Stable GSs correspond to $n_{r}=0$ [panels (a)--(c) and (h)]. In addition,
Fig. \ref{Fig2}(i) displays an example of a loosely bound but also stable GS
supported by a weaker OL, with $s=5$. The parameters are adjusted to $N=450$
atoms [$N=350$ in Fig. \ref{Fig2}(i)]; in Ref. \cite{ober1}, the GS had
$\simeq250$ atoms. The right-hand image in each panel is a perspective image,
while the top and bottom left plots are axial and lateral views.

\begin{figure}[ptb]
\begin{center}
\includegraphics[
width=8.6cm
]{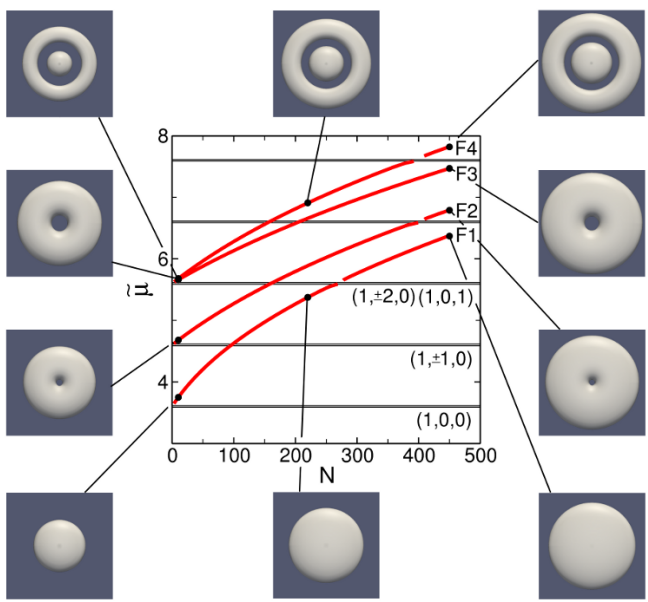}
\end{center}
\caption{(Color online) The $\widetilde{\mu}(N)$ curves representing the four
lowest-lying GS families. The inserted panels display the solitons at marked
points. The field of view in each panel is $4.6\operatorname{\mu m}%
\times4.6\operatorname{\mu m}$, in terms of the $^{87}$Rb condensate.}%
\label{Fig3}%
\end{figure}

As shown in Fig. \ref{Fig3}, GS families are represented by $\widetilde{\mu
}(N)$ curves, each approaching a certain band at $N\rightarrow0$, whose set
$(n,m,n_{r})$ is used to label the families. As $N$ increases, the size of the
GSs increases too, keeping a characteristic soliton structure. The GS with the
lowest chemical potential, of type $(1,0,0)$, which corresponds to point F1 in
Fig. \ref{Fig1}(b), is shown in Fig. \ref{Fig2}(a). GSs of this type are
disk-shaped objects localized within a single cell of the axial OL. While the
radial shape of this GS approaches the ground state of the corresponding HO at
$N\rightarrow0$, the contribution of radially excited states becomes more
pronounced with the increase of $N$, making the GS a fully 3D mode. In
particular, point F1 in Fig. \ref{Fig1}(b), with $\widetilde{\mu}=6.37$,
corresponds to $\mu=7.37\hbar\omega_{\bot}$ in physical units, which is
\emph{much larger} than the radial-excitation quantum, $\hbar\omega_{\bot}$.
The decomposition of this GS over the basis of radial HO modes yields a
mixture of states $n_{r}=0,1,$ and $2$, with respective weights $76\%,22\%,$
and $2\%$

As per azimuthal index $m$, the GSs in the $(1,1,0)$ and $(1,2,0)$ families
carry vorticities $1$ and $2$; see Figs. \ref{Fig2} and \ref{Fig3}. (The
vorticity may also be embedded into solitons in the case of the tight
transverse confinement, with $\hbar\omega_{\perp}\gg E_{R}$ \cite{Luca}, as
well as into GSs supported by the OL in the 2D geometry \cite{vortex}.)\ The
GSs corresponding to $n_{r}\geq1$ feature a complex radial structure, composed
of many HO modes. Nevertheless, these solitons exhibit a set of zero-density
rings reminiscent of the HO wave functions with the respective values of
$n_{r}$; see Figs. \ref{Fig2}(d), \ref{Fig2}(f) and \ref{Fig2}(g) for
$n_{r}=1,2,$ and $3$.

\begin{figure}[ptb]
\begin{center}
\includegraphics[width=8.6cm ]{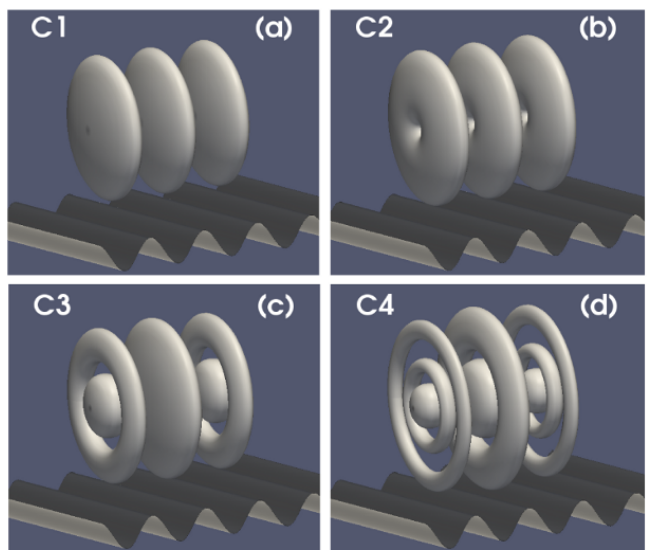}
\end{center}
\caption{(Color online) Gap-soliton complexes corresponding to points C1--C4
in Fig. \ref{Fig1}(b). The underlying 1D lattice potential is also shown.}%
\label{Fig4}%
\end{figure}

The higher-order ($n=2$) band of the axial potential gives rise to GS families
of types $(2,\pm1,0)$. As seen in Fig. \ref{Fig2}(h), they exhibit two major
peaks in the axial direction with a node between them, all squeezed into a
\emph{single} lattice cell (in the 1D GPE, solitons of this type are known as
subfundamental GSs \cite{sub}). On the other hand, because band gaps emerging
at large values of the chemical potential are very narrow, no GS of type
$(2,0,0)$ exists here, as it is not able to place itself within the
corresponding narrow gap. It is worth noting, too, that GSs of the $(1,0,3)$
and $\left(  1,1,2\right)  $ types are found \emph{inside} the $(2,1,0)$ and
$\left(  2,0,0\right)  $ Bloch bands, respectively [the former case
corresponds to point F7 in Figs. \ref{Fig1}(b) and \ref{Fig2}(g)], which is
possible because modes with different azimuthal numbers $m$ do not mix. In
fact, these may be understood as examples of \textquotedblleft embedded
solitons", whose chemical potential falls within linear bands. While they were
studied in detail in 1D models \cite{embedded}, no example of multidimensional
embedded solitons has been reported previously.

Fundamental gap solitons play an important role as elementary building blocks
of higher order localized nonlinear structures \cite{Wu1}. Some examples of
the latter, generated from the symmetric linear combination of three
fundamental solitons, are shown in Fig. \ref{Fig4}. The gap solitons displayed
in Figs. \ref{Fig4}(a) and \ref{Fig4}(b), which are composed of three
identical $(1,0,0)$ and $(1,1,0)$ solitons, respectively, and correspond to
points C1 and C2 in Fig. \ref{Fig1}(b), have exactly the same chemical
potentials as their respective elementary constituents ($\widetilde{\mu}=6.37$
and $6.79$) but contain approximately three times as many particles
($N=1372$). The compound gap solitons in Figs. \ref{Fig4}(c) and
\ref{Fig4}(d), which have the same particle content as the previous ones but
are higher in the spectrum (points C3 and C4), are bound states of fundamental
constituents of different types. The former is composed of one $(1,0,0)$ and
two $(1,0,1)$ solitons, while the latter contains one $(1,0,1)$ and two
$(1,0,2)$ fundamental solitons.

\section{III. STABILITY}

\begin{figure}[ptb]
\begin{center}
\includegraphics[
width=8.6cm
]{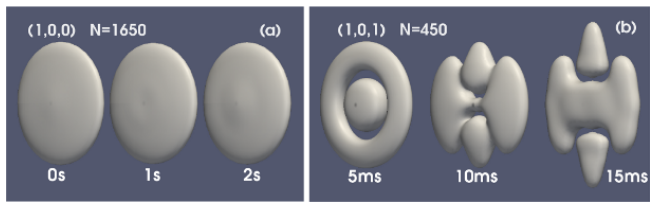}
\end{center}
\caption{(Color online) (a) The stable evolution of a GS of the $(1,0,0)$ type
after the application of the perturbation. (b) The instability of the soliton
of the $\left(  1,0,1\right)  $ type, from Fig. \ref{Fig2}(d).}%
\label{Fig5}%
\end{figure}

The stability of the GSs was tested by simulating their perturbed evolution.
To this end, Eq. (\ref{eq1}) was solved by the Laguerre-Fourier pseudospectral
method, using the third-order Adams-Bashforth time-marching scheme. To include
all potentially dangerous disturbances, we perturbed the GSs by simultaneously
increasing the OL depth and decreasing its period, both by $2\%$, translating
the OL in the axial direction by $2\%$ of the lattice period, and applying a
$2\%$ quadrupole deformation to the transverse trapping potential. After
waiting for $t=1$ $%
\operatorname{ms}%
$, the combined perturbation was removed, allowing the system to evolve for
$2$ $%
\operatorname{s}%
$. The simulations demonstrate that the fundamental GSs of the $(1,0,0)$ and
$(1,1,0)$ types remain stable, except in narrow regions close to edges of the
corresponding band gaps. As a representative example, Fig. \ref{Fig5}(a)
displays the evolution of the soliton of the $(1,0,0)$ type, corresponding to
point F9 in Fig. \ref{Fig1}(b), which represents a GS built of $1650$ atoms
with $\mu=10.3\hbar\omega_{\bot}$. While the latter is much greater than the
transverse quantum $\hbar\omega_{\bot}$, the GS remains stable under the
action of the 3D perturbations. The loosely bound GSs trapped in weaker OLs,
such as the one in Fig. \ref{Fig2}(i), are stable as well.

Although vortices with the $m=2$ are usually unstable against splitting
\cite{PRL06}, they may be stable in trapped configurations \cite{Pu1}. We have
found that the GS of type $(1,2,0)$ is stabilized by making it heavier: While
it is unstable if built of $\lesssim300$ atoms, the one shown in Fig.
\ref{Fig2}(c), which contains $450$ atoms, is robust in the simulations. On
the other hand, GSs with $n_{r}\geq1$ (that is, those featuring the complex
radial structure) are unstable against quadrupole perturbations. In
particular, Fig. \ref{Fig5}(b) displays the instability for the GS of type
$\left(  1,0,1\right)  $, initiated by a very weak ($0.1\%$) quadrupole
deformation of the trapping potential, acting for $0.2$ $%
\operatorname{ms}%
$. GSs with $n=2$ are also unstable, similar to the previously mentioned
\textquotedblleft subfundamental solitons" in 1D models \cite{sub}. Finally,
the stability of GS complexes coincides with that of their fundamental constituents.

\section{IV. CONCLUSION}

Up to now, matter-wave GSs were actually created in shallow lattices
($s\simeq0.7$), with a weak radial confinement ($E_{R}/\hbar\omega_{\bot
}\simeq44$) \cite{ober1}. The respective linear spectrum does not exhibit true
band gaps. This can be checked using the approximate dispersion relation for
the lowest-energy band in the shallow OL, $E(q)/E_{R}=(qd/\pi-1)^{2}%
-\sqrt{4(qd/\pi-1)^{2}+s^{2}/16}$ \cite{mor1}. As follows from this, the width
of the lowest band, $E(1)-E(0)\simeq0.83E_{R}$, is \emph{much larger} than the
gap, $\hbar\omega_{\bot}\simeq0.027E_{R}$, which separates the replicas
corresponding to higher values of azimuthal number $m$, and hence different
Bloch bands overlap in this case. (For parameters of the experiment from Ref.
\cite{ober1}, true gaps open up only at $s\geq16$.) Therefore, the solitary
modes obtained under such conditions are actually quasisolitons, decaying on a
time scale of $\simeq60$ $%
\operatorname{ms}%
$ \cite{ober1}. The present results suggest possibilities for the creation of
robust fundamental and compound GSs, including vortical ones, which place
themselves in true band gaps. Simultaneously, this approach opens
possibilities for creating truly multidimensional matter-wave solitons, which
thus far have been elusive in experiments.

In addition to the BEC, the proposed scheme may be used for the creation of 3D
\textquotedblleft light bullets" of the GS type in self-defocusing optical
media with the anomalous group-velocity dispersion. In that case, $t$ and $z$
in Eq. (\ref{eq1}) play the roles of the transmission distance and reduced
time \cite{review}. The corresponding lattice cannot be created as a material
structure, but it may be induced by a beating wave launched at a different
wavelength, similar to the virtual grating used for the creation of optical
GSs in 1D \cite{virtual}. Thus far, no 3D solitons have been created in
optical media.

\begin{acknowledgments}
V.D. acknowledges financial support from Ministerio de Ciencia e
Innovaci\'{o}n FIS2009-07890 (Spain). The work of B.A.M. was supported, in a
part, by the German-Israel Foundation through grant No. 149/2006.
\end{acknowledgments}


\begin{thebibliography}{99}                                                                                               %


\bibitem {mor1}O. Morsch and M. Oberthaler, Rev. Mod. Phys. \textbf{78}, 179 (2006).

\bibitem {BBB1}B. B. Baizakov, B. A. Malomed and M. Salerno, Europhys. Lett.
\textbf{63}, 642 (2003); Phys. Rev. A \textbf{70}, 053613 (2004); J. Yang and
Z. H. Musslimani, Opt. Lett\textit{.} \textbf{28}, 2094 (2003); J. Yang,
I. Makasyuk, A. Bezryadina, and Z. Chen, Stud. Appl. Math. \textbf{113}, 389
(2004); D. Mihalache \textit{et al.},
Phys. Rev. E \textbf{70}, 055603(R) (2004).

\bibitem {1D}F. Kh. Abdullaev \textit{et al.},
Phys. Rev. A \textbf{64,} 043606 (2001); I. Carrusotto, D. Embriaco, and G. C.
La Rocca, \textit{ibid}. 65, 053611 (2002).

\bibitem {sub}N. K. Efremidis and D. N. Christodoulides, Phys. Rev. A
\textbf{67}, 063608 (2003); T. Mayteevarunyoo and B. A. Malomed,
\textit{ibid}. \textbf{74}, 033616 (2006); J. Cuevas,
B. A. Malomed, P. G. Kevrekidis, and D. J. Frantzeskakis, \textit{ibid}.
\textbf{79}, 053608 (2009).

\bibitem {Konotop}B. B. Baizakov, V. V. Konotop and M. Salerno, J. Phys. B:
At. Mol. Opt. Phys\textit{.} \textbf{35}, 5105 (2002); P. J. Y. Louis,
E. A. Ostrovskaya, C. M. Savage, and Y. S. Kivshar, Phys. Rev. A \textbf{67},
013602 (2003); Z. Shi,
J. Wang, Z. Chen, and J. Yang, Phys. Rev. A \textbf{78}, 063812 (2008).

\bibitem {vortex}E. A. Ostrovskaya and Y. S. Kivshar, Phys. Rev. Lett.
\textbf{93}, 160405 (2004); H. Sakaguchi and B. A. Malomed, J. Phys. B: At.
Mol. Opt. Phys. \textbf{37}, 2225 (2004).

\bibitem {ober1}B. Eiermann \textit{et al.},
Phys. Rev. Lett\textit{.} \textbf{92}, 230401 (2004).

\bibitem {gap-wave-experiment}Th. Anker \textit{et al.},
Phys. Rev. Lett. \textbf{94}, 020403 (2005); T. J. Alexander, E. A.
Ostrovskaya, and Y. S. Kivshar, \textit{ibid}. \textbf{96}, 040401 (2005).

\bibitem {Gora}A. E. Muryshev \textit{et al.},
Phys. Rev. Lett. \textbf{89}, 110401 (2002); L. D. Carr and J. Brand,
\textit{ibid}. \textbf{92}, 040401 (2004); M. Matuszewski,
W. Kr\'{o}likowski, M. Trippenbach, and Y. S. Kivshar , Phys. Rev. A
\textbf{73}, 063621 (2006); A. Mu\~{n}oz Mateo and V. Delgado, Phys. Rev. A
\textbf{75}, 063610 (2007); \textit{ibid}. \textbf{77}, 013617 (2008); Ann.
Phys. \textbf{324}, 709 (2009).

\bibitem {Luca}L. Salasnich, A. Parola, and L. Reatto, Phys. Rev. A
\textbf{65}, 043614 (2002); \textit{ibid}. \textbf{66}, 043603 (2002); L.
Salasnich,
A. Cetoli, B.A. Malomed, and F. Toigo \textit{ibid}. \textbf{75}, 033622
(2007); L. Salasnich, B. A. Malomed, and F. Toigo,
Phys. Rev. A \textbf{76}, 063614 (2007).

\bibitem {TG}B. Paredes \textit{et al.},
Nature \textbf{429}, 277 (2004).

\bibitem {SKA}S. Adhikari and B. A. Malomed, Europhys. Lett. \textbf{79},
50003 (2007); Physica D \textbf{238}, 1402 (2009).

\bibitem {review}B. A. Malomed, D. Mihalache, F. Wise, and L. Torner, J.
Optics B: Quant. Semicl. Opt. \textbf{7}, R53 (2005).

\bibitem {RCG}R. Carretero-Gonz\'{a}lez, D. J. Frantzeskakis, and P. G.
Kevrekidis, Nonlinearity \textbf{21}, R139 (2008).

\bibitem {embedded}J. Yang, B. A. Malomed, and D. J. Kaup, Phys. Rev. Lett.
\textbf{83}, 1958 (1999); A. R. Champneys,
B. A. Malomed, J. Yang, and D. J. Kaup, Physica D \textbf{152-153}, 340
(2001); J. K. Yang, Phys. Rev. Lett \textbf{91}, 143903 (2003); X. S. Wang,
Z. G. Chen, J. D. Wang, and J. K. Yang, , \textit{ibid}. \textbf{99}, 243901 (2007).

\bibitem {Wu1}M. Matuszewski,
E. Infeld, B. A. Malomed, and M. Trippenbach, Phys. Rev. Lett. \textbf{95},
050403 (2005); Y. Zhang and B. Wu, \textit{ibid.} \textbf{102}, 093905 (2009);
J. D. Wang,
J. K. Yang, T. J. Alexander, and Y. S. Kivshar, Phys. Rev. A \textbf{79},
043610 (2009).

\bibitem {PRL06}A. Mu\~{n}oz Mateo and V. Delgado, Phys. Rev. Lett.
\textbf{97}, 180409 (2006); J. A. Huhtam\"{a}ki \textit{et al}., \emph{ibid.}
\textbf{97}, 110406 (2006); K. Gawryluk, M. Brewczyk, and K. Rzazewski, J.
Phys. B: At. Mol. Opt. Phys. \textbf{39}, L225 (2006).

\bibitem {Pu1}H. Pu \textit{et al}., Phys. Rev. A \textbf{59}, 1533 (1999).

\bibitem {virtual}G. Van Simaeys,
S. Coen, M. Haelterman, and S. Trillo, Phys. Rev. Lett. \textbf{92}, 223902 (2004).
\end{thebibliography}
\end{document}